\documentclass[prd,preprint,nofootinbib]{revtex4}
\usepackage{graphicx}
\usepackage{amssymb}
\usepackage{amsmath}

\preprint{IPMU-10-0138}
\preprint{KEK-TH-1391}

\begin{document}

\newcommand{\beq}{\begin{equation}}   
\newcommand{\eeq}{\end{equation}}
\newcommand{\bea}{\begin{eqnarray}}   
\newcommand{\eea}{\end{eqnarray}}
\newcommand{\bear}{\begin{array}}  
\newcommand {\eear}{\end{array}}
\newcommand{\bef}{\begin{figure}}  
\newcommand {\eef}{\end{figure}}
\newcommand{\bec}{\begin{center}}  
\newcommand {\eec}{\end{center}}
\newcommand{\non}{\nonumber}  
\newcommand {\eqn}[1]{\beq {#1}\eeq}
\newcommand{\la}{\left\langle}  
\newcommand{\ra}{\right\rangle}
\newcommand{\ds}{\displaystyle}
\def\SEC#1{Sec.~\ref{#1}}
\def\FIG#1{Fig.~\ref{#1}}
\def\EQ#1{Eq.~(\ref{#1})}
\def\EQS#1{Eqs.~(\ref{#1})}
\def\GEV#1{10^{#1}{\rm\,GeV}}
\def\MEV#1{10^{#1}{\rm\,MeV}}
\def\KEV#1{10^{#1}{\rm\,keV}}
\def\lrf#1#2{ \left(\frac{#1}{#2}\right)}
\def\lrfp#1#2#3{ \left(\frac{#1}{#2} \right)^{#3}}
\newcommand{\phih}{\hat{\phi}}
\newcommand{\phit}{\tilde{\phi}}

%

\title{
 Running Kinetic Inflation
}

\author{
Kazunori Nakayama${}^{(a)}$ and
Fuminobu Takahashi${}^{(b)}$
}

\affiliation{
${}^{(a)}$ Theory Center, KEK, 1-1 Oho, Tsukuba, Ibaraki 305-0801, Japan\\
${}^{(b)}$ Institute for the Physics and Mathematics of the Universe,
University of Tokyo, Chiba 277-8583, Japan
}

\date{\today}

\begin{abstract}
  We study a recently proposed running kinetic inflation model in
  which the inflaton potential becomes flat due to rapid growth of the
  kinetic term at large inflaton field values.  As concrete examples,  we build a variety of chaotic 
  inflation models in supergravity with e.g. quadratic, linear, and fractional-power potentials. 
  The power of the potential generically increases after inflation, and 
  the inflaton is often massless at the potential minimum in the
  supersymmetric limit, which leads to many interesting
  phenomena. First, the light inflaton mass greatly relaxes severe
  thermal and non-thermal gravitino problems. Secondly, the kination
  epoch is naturally present after inflation, which may enhance the
  gravity waves. Thirdly, since the inflaton is light, it is likely
  coupled to the Higgs sector for successful reheating. The inflaton
  and its superpartner, inflatino, may be produced at the
  LHC. Interestingly, the inflatino can be dark matter, if it is the
  lightest supersymmetric particle.
\end{abstract}


\maketitle

\section{Introduction}
\label{sec1}

The inflation solves theoretical problems of the standard big bang
cosmology, and moreover, quantum fluctuations of an inflaton generate
tiny density perturbations, which can account for the seed of the
structure in the current Universe. The recent WMAP
results~\cite{Komatsu:2010fb} have given strong support to the
inflationary paradigm, and the Planck satellite~\cite{:2006uk} will
narrow down the inflation models in coming years.

While there have been proposed many models, it is a non-trivial task
to construct a successful inflation model, partly because inflaton
properties are poorly understood.  It is customary to assume that, in
the simple slow-roll inflation paradigm, an inflaton is a weakly
coupled field and therefore its kinetic term does not significantly
change during and after inflation.  The flat potential necessary for
the inflation to occur can be realized by symmetry. Alternatively,
without a symmetry, the slow-roll inflation may be attained by
fine-tuning the shape of the potential and/or the initial position of
the inflaton. Also there are a variety of inflation models which do
not come under this
classification~\cite{ArmendarizPicon:1999rj,Silverstein:2003hf,ArkaniHamed:2003uz}.

Recently, a new class of inflation models was proposed by one of the
authors (FT)~\cite{Takahashi:2010ky}, in which the kinetic term grows
as the inflaton field, making the effective potential flat.  This
model naturally fits with a high-scale inflation model such as chaotic
inflation~\cite{Linde:1983gd}, in which the inflaton moves over a
Planck scale or even larger within the last $50$
e-foldings~\cite{Lyth:1996im}.  This is because the precise form of
the kinetic term may well change after the inflaton travels such a
long distance.  In some cases, the change could be so rapid, that it
significantly affects the inflaton dynamics.  We call such model as
{\it running kinetic inflation}.  Interestingly, the power of the
inflaton potential generically changes in this class of inflation
models.  For instance, we can build a model that generates a quadratic
potential for the inflaton during inflation and a quartic one after
inflation. Such behavior of the scalar potential has various
cosmological implications.

In this paper we explore phenomenological aspects of the running
kinetic inflation model.  In most cases the inflaton has a mass of
order of the gravitino mass, which makes it easy to avoid the
thermal~\cite{Weinberg:zq,Krauss:1983ik} and
non-thermal~\cite{Kawasaki:2006gs,Endo:2007ih} gravitino problems. This
is because the reheating temperature tends to be low and because the
non-thermal gravitino production can be kinematically suppressed.  We
also estimate the amplitudes of the primordial gravity waves, which
may be enhanced due to a possible kination epoch after inflation, and
we discuss how the reheating proceeds. As we will see later, it is
likely that the inflaton is coupled to the Higgs sector for successful
reheating, and then the inflatino becomes a natural candidate for dark
matter, and can be produced at the LHC.

Let us briefly mention differences from works in the past.  
While our model is motivated by a simple idea that the kinetic term may well
change its form in the large-scale inflation such as the chaotic
inflation, there are stringy inflation models that exhibit the similar
behavior of the kinetic
term~\cite{Silverstein:2008sg,McAllister:2008hb}.  As far as the
inflaton dynamics is concerned, there is no significant difference if
the potential takes the same form.  However, since our model is based
on supergravity, we can unequivocally discuss the supersymmetry (SUSY)
breaking effects, thermal and non-thermal gravitino problems, and
reheating the standard model (SM) sector. In particular, the last
point enables us to discuss the possibility of the inflatino to become
dark matter.  There are references
(e.g. Refs.~\cite{Tashiro:2003qp,Sami:2004xk, Chung:2007vz}) which studied
the enhancement of the primordial gravity waves due to a kination
epoch after inflation.\footnote{In our context, the kination epoch is such that the inflaton 
energy decreases faster than radiation.} However, the kination is usually 
realized by introducing an additional degrees of freedom with an
appropriate property (such as quintessence) or assuming a fine-tuned
scalar potential for the inflaton.  In contrast, the kination is, in a
sense, built-in to the running kinetic inflation, because the power of
the potential generically increases after inflation. Besides, a light
inflaton and its coupling to the Higgs sector has been considered in
e.g. Refs.~\cite{Shaposhnikov:2006xi,Bezrukov:2009yw} assuming the $\lambda \phi^4$
inflation, which is however disfavored by the WMAP data. To avoid a
tension with observation, it is often assumed that the inflaton has a
non-minimal coupling to the gravity~\cite{Salopek:1988qh,Futamase:1987ua,Spokoiny:1984bd,Fakir:1990eg,Komatsu:1997hv}
(see also Refs.~\cite{Einhorn:2009bh,Lee:2010hj,Ferrara:2010yw,Ferrara:2010in,Kallosh:2010ug} 
for the inflation with non-minimal coupling to gravity in supergravity).
This however entails a large coupling, which calls for explanation.\footnote{
It is difficult to quantify the amount of tuning until we understand how a small or large coupling arises, though.
For instance, an extremely small coupling $\sim 10^{-13}$ in the quartic chaotic inflation can arise
from a coupling of $O(10^{-7})$ in the superpotential~\cite{Kasuya:2003iv}.
}  This idea attracted much attention recently since the proposal of the SM Higgs inflation~\cite{Bezrukov:2007ep}, and
the issue of unitarity has been discussed in 
Refs.~\cite{Barbon:2009ya,Burgess:2009ea,Lerner:2009na,Hertzberg:2010dc}.   On the other hand, it
is possible in our model to realize a situation that the scalar
potential during inflation is given by a quadratic potential which is
consistent with observation, while it becomes a quartic one after
inflation, without introducing a large non-minimal coupling to gravity.  Moreover, as we will see below, the inflaton
is assumed to respect a discrete $Z_k$ symmetry, which can be
identified with the one often introduced to solve the $\mu$-problem in
the extension of the minimal supersymmetric standard model (MSSM).
Thus, it is natural also from a symmetry point of view to expect a
coupling to the Higgs sector in our model.

The rest of the paper is organized as follows. In Sec.~\ref{sec2}, we
study in detail the running kinetic inflation model such as the
inflaton dynamics, the mass spectrum, and the reheating processes. In
Sec.~\ref{sec3}, we estimate how much the gravity waves can be
enhanced by the kination epoch after inflation. In Sec.~\ref{sec4} we
briefly point out that our running kinetic inflation naturally fits
the so called the nearly minimal SUSY SM (nMSSM).  The last section is
devoted to conclusions.

\section{Running Kinetic Inflation}
\label{sec2}

\subsection{Basic idea}
Before going to a realistic inflation model, let us first give our
basic idea.  Suppose that a real scalar field $\phi$ has the following
Lagrangian,
\beq
{\cal L} \;=\; \frac{1}{2} \,f(\phi)\, \partial^\mu \phi \partial_\mu \phi - V(\phi),
\label{kin}
\eeq
where the scalar $\phi$ is an inflaton.  We assume that it is
canonically normalized at the potential minimum:
\beq
f(\phi_{\rm min}) \;=\; 1.
\eeq
This does not necessarily mean that $f(\phi)$ remains close to $1$
during inflation, especially if the inflaton moves over some large
scale, e.g., the GUT or Planck scale.  Suppose that the behavior of
$f(\phi)$ can be approximated by $f(\phi) \approx \phi^{2n-2}$ with an
integer $n$ over a certain range of $\phi$ .  Here and in what follows
we adopt the Planck unit, $M_P \simeq 2.4 \times \GEV{18} = 1$, unless
otherwise stated.  Then, the scalar potential $V(\phi)$ would take a
different form, when expressed in terms of the canonically normalized
inflaton field, $\hat{\phi} \equiv \phi^n/n$.  For instance, let us
consider a case of $n=2$.  Then the quadratic (quartic) potential,
$V(\phi) \propto \phi^2 (\phi^4)$, becomes a linear (quadratic) term
$V(\hat{\phi}) \propto \hat{\phi} (\hat{\phi}^2)$.  Thus, such a
strong dependence of the kinetic term on the inflaton field changes
the inflation dynamics dramatically.

Let us consider another example, 
\beq
f(\phi) \;=\; \kappa+ \phi^2
\label{eq:example2}
\eeq
with $0<\kappa \ll 1$. For $\phi \gtrsim \sqrt{\kappa}$, we can
approximate $f(\phi) \approx \phi^2$, and this model is reduced to the
case of $n=2$ in the previous example.  On the other hand, the
canonically normalized field is given by $\hat{\phi} \approx
\sqrt{\kappa} \phi$ for $\phi \ll \sqrt{\kappa}$.  Such behavior of
the kinetic term results in the transition of the scalar potential at
$\phi \approx \sqrt{\kappa}$.  For instance, a mass term becomes a
linear potential for $\phi \gtrsim \sqrt{\kappa}$, and a quadratic
term for $\phi \lesssim \sqrt{\kappa}$:
\beq
V(\phi) = \frac{1}{2} m^2 \phi^2 \;\Longrightarrow \;
V(\hat{\phi})\; \simeq \;\left\{
\bear{cc}
m^2 \hat{\phi} & {\rm~~~ for~~~} \hat{\phi} \gg \kappa \\
\ds{ \frac{m^2}{2\kappa} {\hat \phi}^2} & {\rm~~~ for~~~} \hat{\phi} \ll \kappa 
\eear
\right.
\label{ex2}
\eeq
Thus we can easily realize a linear potential and the transition to a
quadratic term, starting from a mass term and a relatively simple
kinetic term (\ref{eq:example2}).

In principle, if we write the Lagrangian in terms of a canonically
normalized field from the beginning, we can build a model that
exhibits the same dynamics. However, it would then be unclear why the
potential takes such a peculiar form, and it would be non-trivial to
construct a model like (\ref{ex2}).  This is because higher order
terms usually become important at larger field values, and therefore,
one expects that the linear term could be important at small field
values, but it never dominates over the quadratic term at large field
values.  In contrast to the usual notion, the scalar potential in
general becomes flatter at large field values in the running kinetic
inflation.

Lastly we give an extreme example, $f \approx e^{2\phi}$. In this
case, the canonically normalized field is $\hat{\phi} \approx
e^\phi$. The effective scalar potential would be extremely flat for
sufficiently large $\phi$, when expressed in terms of
$\hat{\phi}$. For instance,
\beq
V(\phi) = \frac{1}{2} m^2 \phi^2 \;\Longrightarrow \;
V(\hat{\phi}) =
\frac{1}{2} m^2 \ln^2\hat{\phi}.
\label{ex3}
\eeq
Thus, a large coefficient of the kinetic term is advantageous for
inflation to occur, since the effective potential becomes flat.\footnote{
In Refs.~\cite{Dimopoulos:2003iy,Izawa:2007qa}, it was noted that a large wave function factor 
helps to realize the slow-roll inflation, where the origin of the large factor could be a modulus field.
In our model, the large wave function factor is realized dynamically by the inflaton,
which results in the transition of the inflaton potential, as we will see later. 
}  In
general, however, we would lose predictivity if the growth of the
coefficient $f(\phi)$ is not under control.  As we will see in the
following section, we can impose a certain symmetry which controls the
growth of the kinetic term. Interestingly, the same symmetry also
enables the chaotic inflation in supergravity.

As an immediate consequence of the running kinetic term, we can
construct a variety of inflation models.  For example, it is possible
to realize the inflaton potential which is proportional to $\phi$
during inflation, and to $\phi^4$ after inflation.  This point is
extremely important for both cosmology and phenomenology, and we will
come back to this issue later.

\subsection{Set-up}
Let us build an inflation model in supergravity, in which the kinetic
term grows at large field values in a controlled way. The running
kinetic term naturally occurs if the inflaton moves over more than the
Planck scale during inflation.  This is indeed the case in the chaotic
inflation, which, however, is difficult to implement in supergravity
because of the exponential prefactor $\sim e^K$ in the scalar
potential. We need to introduce some sort of shift symmetry, which
suppresses the exponential growth of the potential.  As we will see
below, the shift symmetry simultaneously fixes the form of the kinetic
term.

\begin{figure}[t]
\includegraphics[scale=0.6]{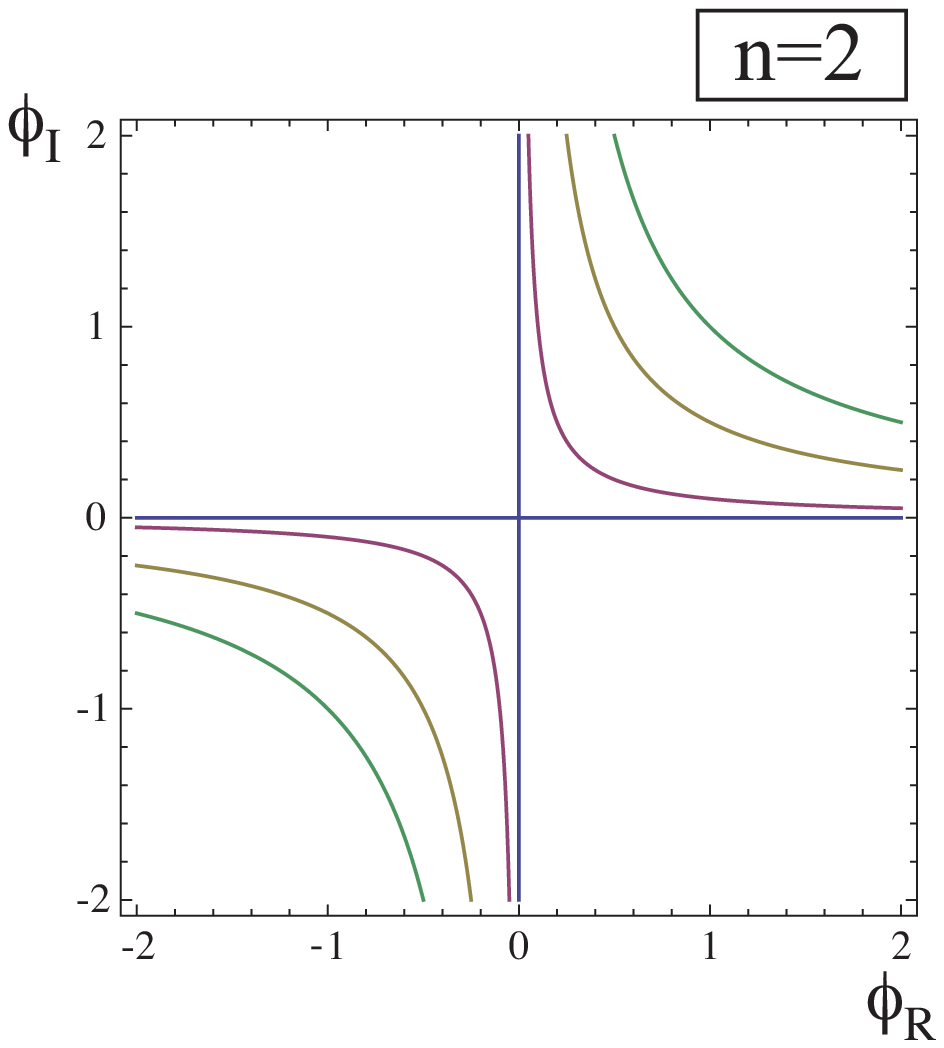}
\includegraphics[scale=0.6]{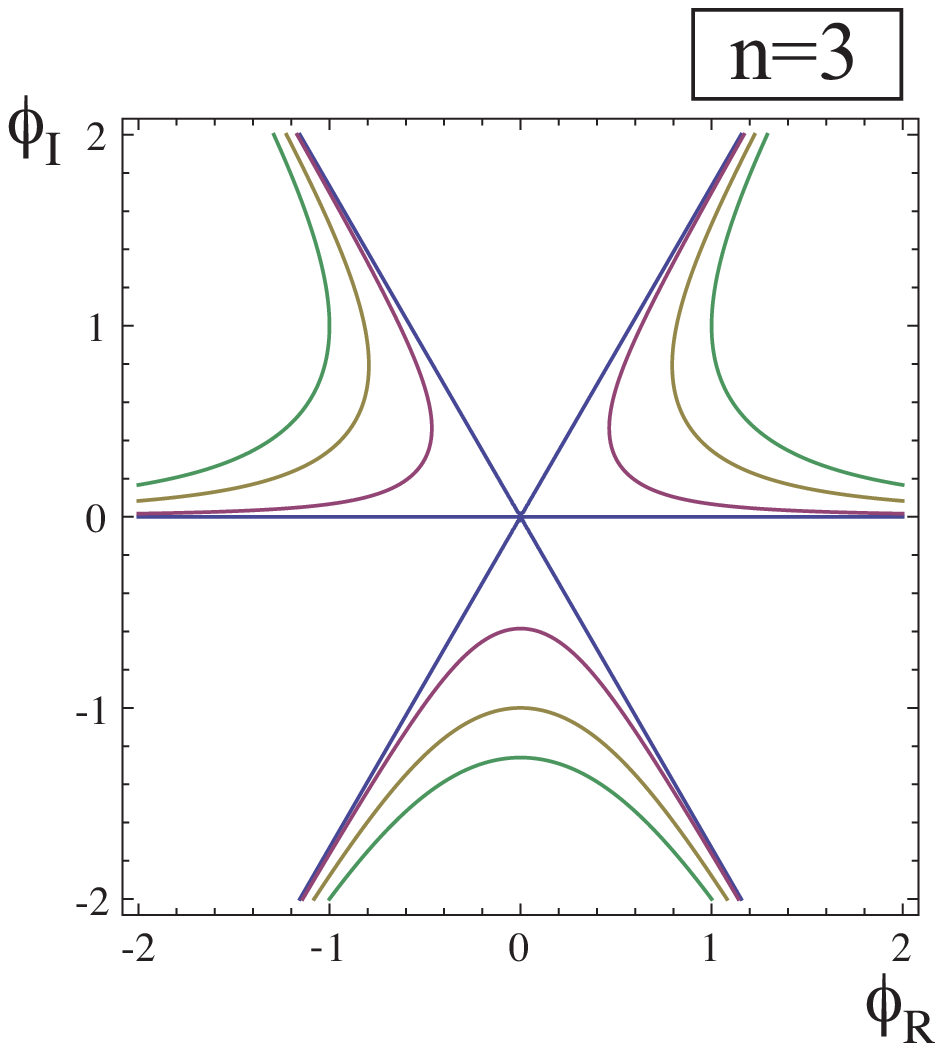}
\caption{ The trajectories generated by the symmetry transformation
  (\ref{sym}) for $n=2$ (left) and $n=3$ (right).  They correspond to
  the contours of ${\rm Im}[\phi^n-\phi^{\dag n}] = 0$, $0.2$, $1$,
  and $2$, respectively.  In the region of $|\phi| \gtrsim 1$, each
  contour corresponds to an inflationary trajectory.  It depends on
  the interactions in the K\"ahler potential which trajectory is
  chosen.  See the text for details.  }
\label{fig:phiRphiI}
\end{figure}

We introduce a chiral superfield, $\phi$, and require that the
K\"ahler potential for $\phi$ is invariant under the following
transformation;
\bea
\phi^n \;\rightarrow\;\phi^n+ \alpha~~~{\rm for~~~}\alpha \in {\bf R}
{\rm ~~~and~~} \phi \ne 0,
\label{sym}
\eea
where $n$ is a positive integer, and $\alpha$ is a transformation
parameter.  The symmetry (\ref{sym}) means that a composite field
${\hat \phi} \sim \phi^n$ transforms under a Nambu-Goldstone like
shift symmetry.  In the case of $n=2$, this is equivalent to imposing
a hyperbolic rotation symmetry (i.e. SO(1,1)) on the real and
imaginary components, $(\phi_R, \phi_I)$, where $\phi =
(\phi_R+i\phi_I)/\sqrt{2}$.  In Fig.~\ref{fig:phiRphiI} we show
trajectories generated by the symmetry transformation (\ref{sym}) for
$n=2$ and $3$.

The K\"ahler potential satisfying the shift symmetry must be a
function of $(\phi^n - \phi^{\dag n})$:
\beq
K\;=\; f(\phi^n-\phi^{\dag n}) = \sum_{\ell=1} \frac{c_\ell}{\ell} \,(\phi^n - \phi^{\dag n})^\ell
\label{Kahler}
\eeq
where $c_\ell$ is a numerical coefficient of $O(1)$ and we normalize
$c_2 \equiv -1$; $c_\ell$ is real (imaginary) for even(odd) $\ell$.
Note that the $|\phi|^2$ term, which usually generates the kinetic
term for $\phi$, is absent if $n \geq 2$. Instead, the kinetic term
arises from the terms of $\ell \geq 2$, whose contribution is
proportional to $(\phi^n-\phi^{\dag n})^{\ell-2}|\phi|^{2n-2}$.  Note
that the lowest component of the K\"ahler potential vanishes for
$\theta = (j/n) \pi$, where $\theta$ is a phase of $\phi$, $\phi =
|\phi| e^{i\theta}$, and $j=0,\cdots,2n-1$.  The presence of such flat
directions along which the K\"ahler potential does not grow is
essential for constructing a chaotic inflation model in supergravity.

We can impose a discrete $Z_k$ symmetry that is consistent with the
shift symmetry (\ref{sym}).  Requiring $(\phi^n-\phi^{\dag n})$, an
invariant under the shift symmetry, be also invariant under the
discrete symmetry up to a phase factor, we find that $k$ must be one
of the divisors of $2n$. If $k=2n$, the $(\phi^n-\phi^{\dag n})$ would
flip its sign, and so, $c_\ell$ with any odd $\ell$ should vanish. If
$k$ is a divisor of $n$, there is no such constraint, since $\phi^n$
itself is invariant under $Z_k$.

In order to have a successful inflation model, let us add two
different breaking terms of the shift symmetry.  First, we add an
interaction in the superpotential:
\beq
W\;=\;\lambda X \phi^m,
\label{intW}
\eeq
where $\lambda$ is a real numerical coefficient to be fixed by the
WMAP normalization of density perturbations. We assume that $X$ and
$\phi$ have U(1)$_R$ charges $2$ and $0$, respectively.\footnote{ It
  is not possible to assign a non-zero $R$-charge to $\phi$, since it
  would be in conflict with the shift symmetry (\ref{sym}), unless a
  discrete $R$ symmetry is considered.  } We assign a $Z_k$ charge to
$X$ and $\phi$ as shown in Table~\ref{charge} in order to suppress
couplings that would spoil the inflation; e.g. $\int d^2\theta\, X$
and $\int d^2\theta\, X^2$ are forbidden.  We require $m \not\equiv 0
~({\rm mod} \,k)$ for this purpose.  As we will see later, $X$ can be
stabilized at the origin during inflation, which suppresses a negative
contribution $-3|W|^2$ in the scalar potential (\ref{sugraV}).  The
interaction (\ref{intW}) then generates a potential $\sim \lambda^2
|\phi|^{2m}$.

\begin{table}[t]
\begin{center}
\begin{tabular}{c||c|c}
&~$\phi$~ &$X$ \\ \hline
 U(1)$_R$&$0$&$2$ \\ \hline
 $Z_k$ & $1$ & $-m$ 
\end{tabular}
\end{center}
\caption{The charge assignment of $\phi$ and $X$.}
\label{charge}
\end{table}

Secondly, let us add a shift-symmetry breaking term,
\beq
\Delta K\;=\; \kappa |\phi|^2,
\label{intK}
\eeq
to cure the singular behavior of the K\"ahler metric at the
origin. Here $0<\kappa \ll 1$ is a real numerical coefficient, and the
smallness is natural in the 't Hooft's sense~\cite{natural}.  This
term is irrelevant during inflation, but it gives a dominant
contribution to the kinetic term some time after inflation. Note that,
once the shift symmetry is broken in the superpotential by
(\ref{intW}), symmetry breaking terms are radiatively generated in the
K\"ahler potential. Indeed, if $m=j n + 1$ for $j=0,1,2,\cdots$, the
breaking term (\ref{intK}) with $\kappa = O(\lambda^2)$ will be
generated as radiative corrections.\footnote{Taking account of the
  WMAP constraint, only $j=0$ and $1$ are allowed.} In this case,
there is essentially one free parameter, $\kappa \sim \lambda^2$.
Also, a coupling of $\phi$ to the standard model (SM) sector could
induce the term (\ref{intK}) similarly, and then $\kappa$ and
$\lambda$ are not related to each other.  In the following we consider
a general case and treat $\kappa$ and $\lambda$ independent
parameters, but we sometimes set $\kappa = \lambda^2$ as a reference
value.

There could be other interactions which violate the shift symmetry,
but throughout this letter we assume that those symmetry breaking
terms are {\it soft} in a sense that the shift symmetry remains a good
symmetry at large $\phi$.  Also, the breaking term (\ref{intK}) is
most important at low energy, and so, we focus on its effect in this
paper.\footnote{ Other symmetry breaking terms could change how the
  transition occurs, but neither the potential during inflation nor
  the inflaton mass at the minimum are modified.  One interesting
  possibility, however, is the non-Gaussianity; see
  Ref.~\cite{Hannestad:2009yx}.  }

The kinetic term of the scalar field is given by
\beq {\cal L}_K \;=\; \left(\kappa - \sum_{\ell=2} \,n^2\,
  c_\ell\,(\ell-1) (\phi^n-\phi^{\dag n})^{\ell-2}|\phi|^{2n-2}
\right) \partial^\mu \phi^\dag \partial_\mu \phi.
\eeq
Note that the second term contains a factor of $(\phi^n-\phi^{\dag
  n})^{\ell-2}$.  In fact, $\phi^n-\phi^{\dag n}$ is constant along
the inflationary trajectory.  Then, the above kinetic term is
equivalent to the second example (\ref{eq:example2}) in the previous
section.  The novelty here is the presence of the shift symmetry
(\ref{sym}), which determines the form of the kinetic term; namely,
$\hat{\phi} \sim \phi^n$ is the dynamical variable at high energies.

The inflaton field value exceeds the Planck scale during inflation,
$|\phi| > 1$.  Since the scalar potential has an exponential prefactor
$\sim e^K$, the inflationary trajectory should be such that $K$ is
minimized. To be explicit, let us define $\eta \equiv
\phi^n-\phi^{\dag n}$ so that the K\"ahler potential is a function of
$\eta$. (Here we neglect the symmetry breaking term (\ref{intK}),
which is not relevant during inflation.) Along the inflationary
trajectory, $\eta$ is stabilized at $\eta_{\rm min}$ satisfying
$f^\prime(\eta)|_{\eta = \eta_{\rm min}} \approx 0$.  This condition
actually means that $\eta$ should be at an extremum. We assume that it
is a local minimum, for the stability of the inflationary
trajectory. As long as $\eta$ remains at the minimum, the terms with
$\ell \geq 3$ does not change the form of the kinetic term
significantly, and so, we focus on the terms with $\ell \leq 2$ in the
following analysis.  Then $\eta$ is stabilized at
\beq
\eta_{\rm min} \;\simeq\; c_1~~~~~{\rm (during~\,inflation)}.
\eeq
The kinetic term  becomes
\beq
{\cal L}_K \;=\; \left(\kappa + n^2 |\phi|^{2n-2} \right) \partial^\mu \phi^\dag \partial_\mu \phi,
\label{kinetic}
\eeq
where we have substituted $c_2 = -1$. After inflation ends at $|\phi|
\sim 1$, the terms with $\ell \geq 3$ becomes irrelevant for the
dynamics. Therefore Eq.~(\ref{kinetic}) is valid both during and after
inflation. Note that, while $\eta$ is stabilized at $\eta_{\rm min}$
for $|\phi| \gtrsim 1$, it is unconstrained for $|\phi| \lesssim
1$. In other words, the inflation is effectively driven by a single
field orthogonal to $\eta$, while the inflaton dynamics after
inflation is described by a complex scalar.

\begin{figure}[t]
\includegraphics[scale=0.7]{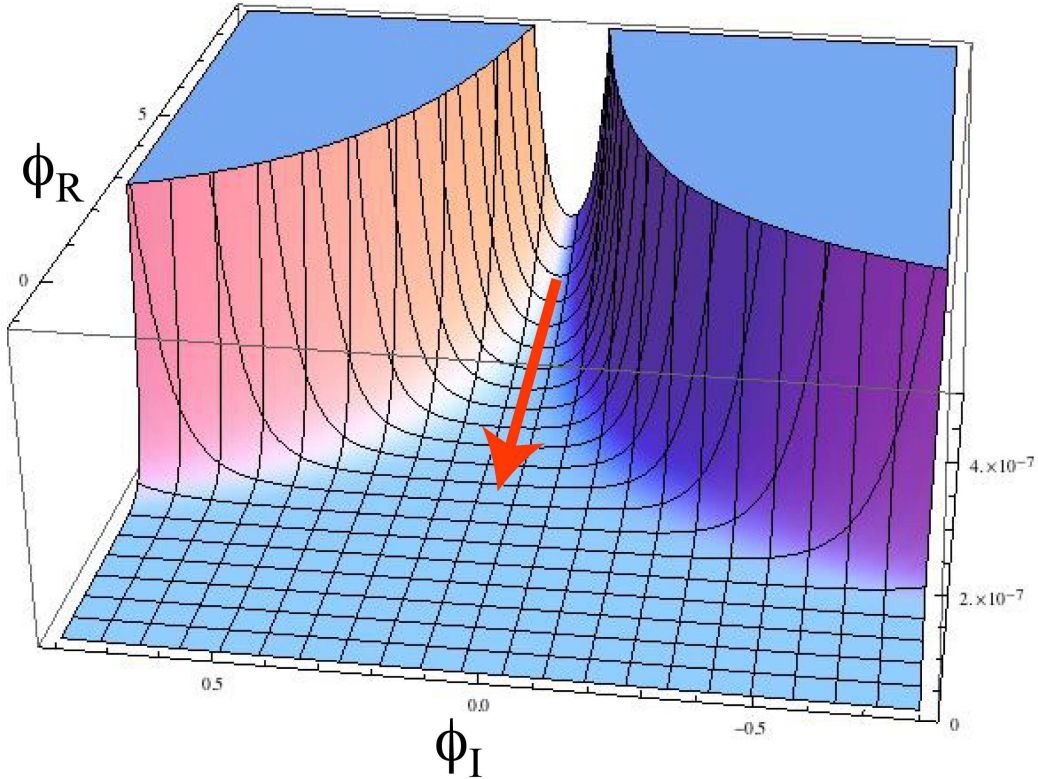}
\caption{ The bird's-eye view of the scalar potential $V(\phi)$ given
  by Eq.~(\ref{V}) with $n=m=2$ and $c_1=i$. The flat direction
  corresponding to the inflationary trajectory is indicated by the
  red arrow. }
\label{fig:Vphi}
\end{figure}

To summarize this section, let us write the K\"ahler and
super-potenetials for the running kinetic inflation:
\bea
\label{Kand W1}
K&=&\kappa|\phi|^2 + c_1 (\phi^n-\phi^{\dag n}) - \frac{1}{2}  (\phi^n-\phi^{\dag n})^2 + |X|^2,\\
W&=&\lambda X \phi^m,
\label{Kand W2}
\eea
where we dropped the terms with $\ell \geq 3$. If we impose the
$Z_{2n}$ symmetry, $c_1$ should vanish.  The scalar potential in
supergravity is given by
\bea
V&=& e^K \left(D_i W K^{i\bar{j}} (D_j W)^* - 3 |W|^2 \right).
\label{sugraV}
\eea
Using the scalar potential, we can show that, during and after
inflation, $X$ has a mass of the order of the Hubble parameter, and
therefore stabilized at the origin.\footnote{The interaction, $K
  \supset -c_X |X|^4$ with $c_X = O(1)$, also gives a Hubble-induced
  mass for $X$.}  Then the relevant Lagrangian for the inflation is
given by
\bea
\label{lagrangian}
{\cal L}&=&  \left(\kappa + n^2 |\phi|^{2n-2} \right) \partial^\mu \phi^\dag \partial_\mu \phi - V,\\
V(\phi)
&\approx&\left\{
\bear{cc}
e^{\kappa |\phi|^2- \frac{c_1^2}{2}}\, \lambda^2 |\phi|^{2m} & {\rm ~~~~~~~~~~~~~~~~~for~~~} |\phi| \gtrsim 1 {\rm ~~and~~} \eta = c_1\\
\lambda^2 |\phi|^{2m}  & {\rm for~~~} |\phi| \lesssim 1 \\
\eear
\right..
\label{V}
\eea
Since we explicitly breaks the shift symmetry (\ref{sym}) by the
$\kappa$ term (\ref{intK}), there appears a non-vanishing exponential
prefactor. However, for $|\phi| < 1/\sqrt{\kappa}$, the exponential
prefactor is of $O(1)$, and therefore can be dropped. Note that the
inflaton does slow-roll even if the exponential prefactor gives a main
contribution to the tilt of the potential.  In Fig.~\ref{fig:Vphi} we
show the shape of the potential $V(\phi)$ with $c_1=1$ in the case of
$n=m=2$.  The flat direction corresponding to the inflationary
trajectory is indicated by the red arrow.  One can see that the
$\eta$, orthogonal to the red arrow, has a large mass during
inflation and the inflationary dynamics is effectively described by a
single field.

\subsection{Inflaton dynamics}
Now we study the inflationary dynamics based on the Lagrangian
(\ref{lagrangian}).  During inflation, $\eta$ is stabilized at
$\eta_{\rm min}$.  So, let us rewrite the Lagrangian in terms of the
inflaton orthogonal to $\eta$.

For $1 < |\phi| \ll \kappa^{-1/2}$, the Lagrangian can be approximated
by
\bea
{\cal L}&\approx&  n^2 |\phi|^{2n-2} \partial^\mu \phi^\dag \partial_\mu \phi - e^{- \frac{c_1^2}{2}} \lambda^2 |\phi|^{2m},\\
&=& \partial^\mu \phih^\dag \partial_\mu \phih - \hat{\lambda}^2 |\phih|^{2m/n},
\eea
where we have defined $\phih \equiv \phi^n$ and $\hat{\lambda} \equiv
e^{- \frac{c_1^2}{4}} \lambda$.  The condition $\eta= \phih-\phih^\dag
= c_1$ is satisfied for
\beq \phih \;= \; \frac{\varphi}{\sqrt{2}} + \frac{c_1}{2}, \eeq where
$\varphi$ is a real scalar. (Remember that $c_1$ is a pure imaginary
constant.)  The Lagrangian for the canonically normalized inflaton is
therefore given by
\beq
\label{V_can}
{\cal L}\; \approx \; \frac{1}{2} \partial^\mu \varphi \partial_\mu \varphi - \hat{\lambda}^2 \lrfp{\varphi}{\sqrt{2}}{2m/n}
\eeq
for $1 < \varphi \ll \kappa^{-n/2}$. Thus, thanks to the shift
symmetry, the inflaton $\varphi$ can take a value greater than the
Planck scale, and the chaotic inflation takes place.

\begin{figure}[t]
\includegraphics[scale=0.6]{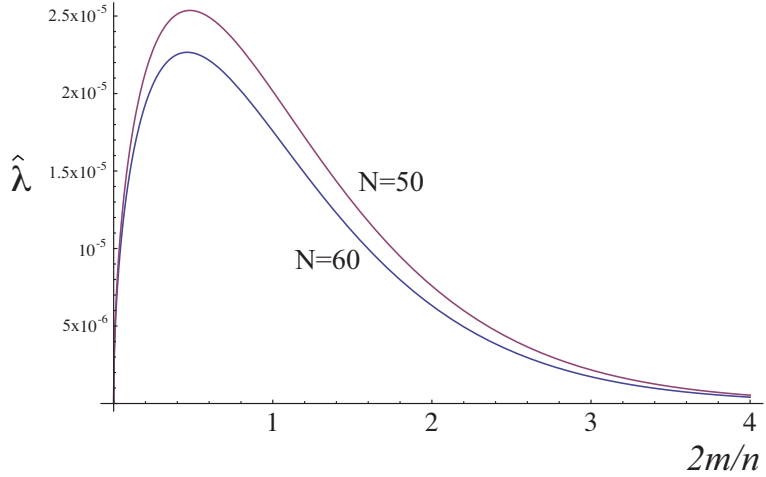}
\caption{
The coupling ${\hat \lambda}$, which accounts for the observed density perturbtion,
as a function of $2m/n$ for the e-foldings $N = 50$ and $60$.
}
\label{fig:lambda}
\end{figure}

The inflaton field durning inflation is related to the e-folding number $N$ as
\beq
\varphi_N \;\simeq\;\sqrt{\frac{4m}{n}N},
\eeq
and the inflation ends at $\varphi = O(1)$ for $m/n = O(1)$. The power
spectrum of the density perturbation is given by
\beq
\Delta_{\cal R}^2 \;\simeq\; \frac{V^{3}}{12 \pi^2 V^{\prime 2}} = (2.43 \pm 0.11)\times 10^{-9},
\eeq
where we have used in the second equality the WMAP
result~\cite{Komatsu:2010fb}. The coupling ${\hat \lambda}$ is
therefore determined as
\beq
{\hat \lambda}\;\simeq\;5.4 \times 10^{-4} \left(2^{-\frac{m}{2n}} \lrfp{m}{n}{\frac{n-m}{2n}} N^{-\frac{m+n}{2n}}\right).
\eeq
We show in Fig.~\ref{fig:lambda} the ${\hat \lambda}$ as a function of
$2m/n$.  We can reproduce the known results for the quadratic and
quartic potentials.  For the quadratic potential $V= (1/2) m^2
\varphi^2$ with $2m/n=2$, the inflaton mass during inflation is given
by $m={\hat \lambda} \simeq 2 \times \GEV{13}$, where we used $N=50$.
Similarly, for the quartic potential $V= ({\tilde \lambda}/4)
\varphi^4$ with $2m/n=4$, we obtain ${\tilde \lambda} = {\hat
  \lambda}^2 \simeq 3 \times 10^{-13}$ for $N=50$.

In order for the inflation driven by (\ref{V_can}) to last for $N$
e-foldings, the following inequality must be met;
\beq
\varphi_N \;\lesssim\;\kappa^{-n/2} ~~\Longleftrightarrow~~~ \kappa \;\lesssim\; \lrfp{n}{4mN}{\frac{1}{n}}.
\eeq
For instance, $\kappa$ must be smaller than $0.1$ for for $n=2$, $m=1$
and $N=50$~\cite{Takahashi:2010ky}.  The spectral index $n_s$ and the
tensor-to-scalar ratio $r$ are respectively given by
\bea
n_s &=& 1-\left(1+\frac{m}{n}\right) \frac{1}{N},\\
r &=& \frac{8m}{n} \frac{1}{N},
\eea
 and they are related to each other by 
\beq
1-n_s \;=\; \frac{n+m}{8m} r.
\eeq

\begin{figure}[t]
\includegraphics[scale=0.6]{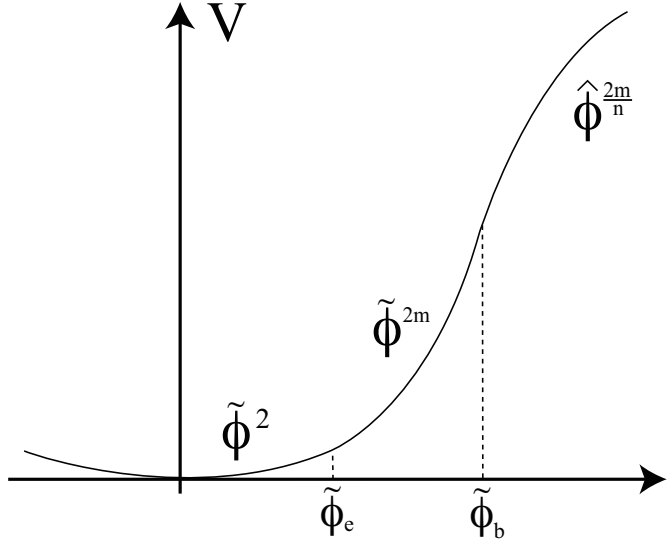}
\caption{ The scalar potential near the origin for $m\geq 2$ with
  respect to the canonically normalized fields $\tilde{\phi}$ and
  $\hat{\phi}$.  The mass term about the origin represents the soft
  SUSY breaking mass.  We define $\tilde \phi_{\rm b} =
  \kappa^{n/(2n-2)}$ and $\tilde \phi_{\rm e} =
  (m_{3/2}\kappa^{m/2}/\lambda)^{1/(m-1)}$.  }
\label{fig:potential_about_origin}
\end{figure}

The inflation ends when the slow-roll condition is violated at
$\varphi \sim 1$, and the inflaton starts to oscillate about the
origin. The $\eta$ is no longer stabilized at $\eta_{\rm min}$ for
$|\phih| < 1$. As the amplitude of the inflaton decreases, the
symmetry breaking term (\ref{intK}) (or the first term in
(\ref{lagrangian})) becomes more important. For $|\phi| <
(\kappa/n^2)^{1/(2n-2)}$, the Lagrangian becomes
\beq
{\cal L}\;\approx\;  \partial^\mu \phit^\dag \partial_\mu \phit- \frac{\lambda^2}{\kappa^m} |\phit|^{2m},
\label{L_at_lowE}
\eeq
where we have defined a canonically normalized field at low scales,
$\phit \equiv \sqrt{\kappa} \phi$.  The power of the scalar potential
changes from $2m/n$ to $2m$ after inflation.  The shape of the
potential for $m \geq 2$ is shown in
Fig.~\ref{fig:potential_about_origin}.  Except for $n=1$, the
potential becomes steeper after inflation. (Note that the model with
$n=m=1$ is equivalent to the model of Ref.~\cite{Kawasaki:2006gs}.)
This is one of the features of the running kinetic inflation.  In
Ref.~\cite{Takahashi:2010ky}, the case of $n=2$ and $m=1$ was mainly
studied, where the potential is linear during inflation and becomes a
quadratic term after inflation.

The inflaton has a large mass at the potential minimum, $m_\phi \simeq
\lambda/\sqrt{\kappa}$, only if $m=1$. On the other hand, for $m \geq
2$, the inflaton remains massless in the SUSY limit.  This opens up a
very interesting phenomenon; for instance, if take $n=m \geq 2$, the
inflaton is massive during inflation, but it becomes massless after
inflation!  Thus, the inflaton decay process will be significantly
modified compared to the case of the massive inflaton.  In the next
section we will study the inflaton mass at the potential minimum,
taking account of the SUSY breaking effect.

\subsection{Inflaton mass spectrum}
\label{ims}
One of the features of the running kinetic inflation model is that the
scalar potential becomes steeper some time after inflation. In the
SUSY limit, the inflaton mass at the potential minimum depends on the
value of $m$. If $m=1$, both inflaton and inflatino have a large SUSY
mass, while they are massless for $m\geq 2$. In the latter case, once
the SUSY breaking effect is taken into consideration, the inflaton
and/or inflatino acquire a soft SUSY breaking mass that depends on
$n$. If $n=2$, both the inflaton and inflatino acquire a soft mass
heavier than the gravitino mass, while the inflaton mass is of order
of the gravitino mass and the inflatino remains massless for $n\geq
3$. We consider these three cases.

\subsubsection{A case of  $m=1$}
Both the inflaton and $X$ have a large and degenerate SUSY mass
$\simeq \lambda/\sqrt{\kappa}$, and these two fields are maximally
mixed, once a constant term in the superpotential is taken into
account~\cite{Kawasaki:2006gs}.  Depending on the precise value of
$\kappa$, the mass can vary from $\GEV{14}$ up to the Planck mass.
Since the inflaton mass is so large that the SUSY breaking effect is
negligible.

\subsubsection{A case of $n=2$ and $m\geq 2$}
In the case of $n=2$ and $m\geq 2$, the inflaton is massless at the
potential minimum in the SUSY limit.  However, the $\ell=1$ term gives
a soft SUSY breaking mass much larger than the gravitino mass.

Let us consider the case of $n=m=2$, and the case of $m \geq 3$ can be
treated in the same way.  The effective potential around the origin is
proportional to $|\phit|^4$ (see Eq.~(\ref{L_at_lowE})). Therefore the
inflaton as well as $X$ remain massless in the SUSY limit. In order to
estimate their masses we need to take account of the SUSY breaking
effect.

To this end, let us introduce a SUSY breaking superfield $Z$, which
has a non-zero F-term, $\la F_Z \ra \equiv \mu^2 = \sqrt{3}m_{3/2}$,
where $m_{3/2}$ is the gravitino mass.  We fix $\la Z \ra = 0$, and do
not consider its dynamics in the following analysis.  The K\"ahler and
super-potentials are
\bea
K &=& \kappa |\phi|^2 +  c_1(\phi^2 - \phi^{\dag 2}) - \frac{1}{2} (\phi^2 - \phi^{\dag 2})^2 + |X|^2 + |Z|^2+\cdots,\\
W &=& \lambda X \phi^2 + \mu^2 Z + W_0,
\label{KW22}
\eea
where $W_0 = m_{3/2}$. In general there could be couplings among $Z$,
$\phi$ and $X$ in the K\"ahler potential, which are denoted by the
dots.  As long as they are suppressed by the Planck scale, the
following result does not change significantly, and so, we drop them
in the following.

Let us rewrite the K\"ahler and super-potentials in terms of $\phit$:
\bea
K &=& |\phit|^2 +  \frac{c_1}{\kappa}(\phit^2 - \phit^{\dag 2}) - \frac{1}{2\kappa^2} (\phit^2 - \phit^{\dag 2})^2 + |X|^2 + |Z|^2,\\
W &=& \frac{\lambda}{\kappa} X \phit^2 + \mu^2 Z + W_0.
\label{KW}
\eea
Using the supergravity scalar potential (\ref{sugraV}), one can show
that $\phit$ and $X$ are stabilized at the origin.  Their masses at
the origin are
\bea
\label{minf_case2}
m_{\phit} &=& \sqrt{1+\frac{4|c_1|^2}{\kappa^2}} m_{3/2} \;\approx\; \frac{2 |c_1|}{\kappa} m_{3/2},\\
m_X&=& m_{3/2},
\eea
where we have used $\kappa \ll 1$ and $|c_1| = O(1)$.  As long as the
inflaton is coupled to the SUSY breaking field only gravitationally,
the inflaton is stabilized at the origin. This is not necessarily the
case for $X$; the origin could be destabilized depending on a possible
quartic coupling $|X|^2|Z|^2$ in the K\"ahler potential.  Note that
$\phit$ and $X$ do not mix in contrast to the case of $n=m=1$.  The
inflatino has the same mass as the inflaton.

The result changes if we impose the $Z_{2n}$ discrete symmetry. In this case
$c_1$ should vanish, and the inflaton mass becomes of order of the gravitino mass.

\subsubsection{A case of $n \geq 3$ and $m \geq 2$}
Similarly we consider the case of $n=m=3$; the other cases can be
treated along the same line.  In the case of $n=m=3$, the inflaton and
$X$ are massless in the SUSY limit as in the previous case.  Let us
write down the K\"ahler and super-potentials (\ref{Kand W1}) and
(\ref{Kand W2}) in terms of $\phit$:
\bea
K &=& |\phit|^2 +  \frac{c_1}{\kappa^{3/2}}(\phit^3 - \phit^{\dag 3}) - \frac{1}{2\kappa^3} (\phit^3 - \phit^{\dag 3})^2 + |X|^2 + |Z|^2,\\
W &=& \frac{\lambda}{\kappa^{3/2}} X \phit^3 + \mu^2 Z + W_0.
\label{KW22-can}
\eea
Using the supergravity scalar potential (\ref{sugraV}), one can show
that $\phit$ and $X$ are stabilized at the origin.  Their masses at
the origin are
\bea
m_{\phit}&=& m_X= m_{3/2}.
\eea
For a general K\"ahler potential, both the inflaton and $X$ have a
mass of $O(m_{3/2})$.  The inflatino still remains massless. As we
will see later, if we couple the inflaton to the Higgs sector, the
inflatino acquires a mass after the electroweak phase transition.

\subsection{Solution to the non-thermal gravitino production}

In order to have a successful inflation model, one needs to recover
the hot radiation-dominated Universe after inflation. This is achieved
by the inflaton decay, which transfers the inflaton energy to
radiation. Recently it was pointed out that a significant amount of
gravitinos can be produced by the inflaton
decay~\cite{Kawasaki:2006gs}; the resultant gravitino abundance is
given by an increasing function of the inflaton mass and VEV at the
potential minimum, giving a preference to a light mass and a small
VEV.

Our running kinetic inflation model is a variant of the chaotic
inflation, and so, it belongs to a high-scale
inflation. Interestingly, our model avoids the gravitino
overproduction problem in two ways.  First, the inflaton has a
discrete symmetry, which forbids the inflaton VEV. Even if the
symmetry is explicitly broken by a small amount, the induced VEV could
be suppressed. Secondly, the inflaton is very light at the minimum for
$m \geq 2$, and indeed, it is massless in the SUSY limit. As we have
seen before, the inflaton mass is of order of the gravitino mass in
most cases, which can suppress the inflaton decay into the gravitinos.
Such a kinematic suppression of the gravitino production is a new and
attractive solution to the non-thermal gravitino production.  What is
particularly nice about our model is that the inflaton potential
during inflation is different from that at the minimum, and the
inflation model can be well within the WMAP allowed region, without
invoking a non-minimal coupling to the gravity.

\subsection{Reheating}
\subsubsection{A case of $m=1$}
In order to reheat the Universe, the inflaton must have some couplings
with the SM sector.  In the case of $m=1$, the inflaton mass is
heavier than $\GEV{14}$, and the inflaton and $X$ are maximally mixed
in general~\cite{Kawasaki:2006gs}.  Therefore $\phi$ can decay into
the SM particles through couplings of $X$ with the SM sector. For
instance, if we introduce the coupling with Higgs
doublets~\cite{Kawasaki:2000yn},
\beq
W \;=\; \lambda_X X H_u H_d,
\eeq
where $\lambda_X$ is a numerical coefficient and $H_{u(d)}$ is the
up(down)-type Higgs doublet, the reheating temperature will become
\beq
T_R \;\sim\;\GEV{10} \lrf{\lambda_X}{10^{-5}} \lrfp{m_\phi}{\GEV{14}}{1/2}.
\eeq
Here we have assumed that $H_u H_d$ has a $R$-charge $0$ and a
negative parity under the $Z_2$ symmetry. Alternatively, if we allow a
symmetry-breaking term $\delta (\phi+\phi^\dag) = \delta/\sqrt{\kappa}
(\hat{\phi}+\hat{\phi}^\dag)$ in the K\"ahler potential, the inflaton
decays into the SM particles through the gravitational couplings with
the top Yukawa interaction and the SU(3)$_C$ gauge
sector~\cite{Endo:2006qk,Endo:2007ih}. The reheating temperature will
become
\beq
T_R \;\sim\; 5 \times \GEV{6} \lrf{\delta/\sqrt{\kappa}}{10^{-3}} \lrfp{m_\phi}{\GEV{14}}{3/2}.
\eeq
Note that the $\delta$-term violates both the shift and $Z_2$
symmetries.  If the inflaton mass $m_\phi$ is about $\GEV{16}$, the
reheating temperature becomes about $\GEV{10}$, which is high enough
for the thermal leptogenesis to
work~\cite{Fukugita:1986hr,Endo:2006nj}.  However, it is in general
difficult to satisfy the constraints from the non-thermal gravitino
problem when the inflaton decay is induced by the gravitationally
suppressed coupling, unless the gravitino is either extremely light
$m_{3/2} \leq O(10)$\,eV~\cite{Viel:2005qj} or very heavy $m_{3/2}
\geq O(10^2)$\,TeV {\it and} the R-parity is broken. We also note that
the inflaton has an approximate U(1) symmetry at the origin, and
Q-balls~\cite{Coleman:1985ki} may be formed; however, the charge is
relatively small and so it does not affect the reheating
process.\footnote{FT thanks M.~Kawasaki and S.~Kasuya for useful
  discussion.}

\subsubsection{A case of $n=2$ and $m\geq 2$}
In this case, both the inflaton and $X$ are massless in the SUSY
limit. However, once the SUSY breaking is taken into account, the
inflaton becomes heavier than the gravitino, while the mass of $X$ is
of order of the gravitino mass. In contrast to the previous case,
there is no significant mixing between $\phi$ and $X$. So, we need to
couple $\phi$ to the SM sector. For instance, we may couple $\phi$ to
the Higgs sector,
\beq
W\;=\; \lambda_\phi \phi H_u H_d,
\label{phi-higgs}
\eeq
where we have assigned an R-charge $2$ and the $Z_k$ charge $-1$ to
$H_u H_d$. The numerical coefficient $\lambda_\phi$ breaks the shift
symmetry, and so, we may assume $\lambda_\phi \sim \lambda$, although
not necessarily.  Assuming the inflaton mass is much heavier than the
Higgs and Higgsinos, the resultant decay temperature would be
\beq
T_R \;\sim\; \GEV{7}\, |c_1|^{1/2} \lrf{\lambda_\phi}{10^{-5}} \lrfp{\kappa}{10^{-4}}{-\frac{1}{2}} \lrfp{m_{3/2}}{{\rm TeV}}{\frac{1}{2}},
\label{TR_case2}
\eeq
where we have used (\ref{minf_case2}).  Note that, for the parameters
chosen in the above expression, the reheating temperature is
comparable to the inflaton mass. If the reheating temperature
calculated by equating the decay rate and the Hubble parameter exceeds
the inflaton mass, the decay process will be significantly affected by
the back reaction. This is because, when the decay products acquire a
thermal mass comparable to the inflaton mass, the available phase
space will be reduced~\cite{Kolb:2003ke}. That is, the reheating
temperature cannot exceed the inflaton mass, as long as it proceeds
through a perturbative decay. As a result, the reheating temperature
is given by the inflaton mass, if the naively calculated reheating
temperature turns out to be larger than the inflaton mass:
\beq
T_R \;\sim\; m_\phi = \frac{2 |c_1|}{\kappa} m_{3/2},
\label{eq:tr_upper}
\eeq
which is used if $T_R$ given by Eq.~(\ref{TR_case2}) exceeds $m_\phi$.
The validity of (\ref{eq:tr_upper}) will be discussed in the next section.

\subsubsection{A case of $n\geq 3$ and $m\geq 2$}
The main difference from the previous case is that the inflaton mass
is of order of the gravitino mass, and the reheating temperature
becomes lower, accordingly. In particular, in order to reheat the
Universe with such a light inflaton mass, the inflaton must have
unsuppressed couplings to the SM sector.  If it has only
gravitationally suppressed interactions, there would be cosmological
difficulties; the reheating temperature is too low to be consistent
with BBN, and the LSPs may be overproduced by the inflaton
decay~\cite{Endo:2006ix}, as long as we consider $m_{3/2} \lesssim
O(100)$TeV.  Thus, we consider a possibility to couple the inflaton to
the SM particles at the renormalizable level.  The only possibility is
to couple $\phi$ to the Higgs sector as in Eq.~({\ref{phi-higgs}).  If
  we calculate the reheating temperature by equating the decay rate to
  the Hubble parameter, we would obtain a temperature much higher than
  the inflaton mass.  Thus, according to the above argument, the
  reheating temperature is equal to the inflaton mass:
\beq
T_R \;\sim m_\phi \sim m_{3/2}.
\eeq
As a corollary, the inflaton mass, therefore the gravitino mass, needs
to be heavier than $O(10)$\,MeV to be consistent with
BBN~\cite{Kawasaki:1999na,Hannestad:2004px,Ichikawa:2005vw}, since
otherwise the reheating would not be completed before BBN.

Let us discuss the validity of the reheating temperature, $T_R \sim
m_{\phi}$.  So far we have not taken into account that, before the
inflaton decay is completed, the decay products, $H_u$ and $H_d$, have
masses of $O(\lambda_\phi |\phi|)$, which may prevent the inflaton
decay to proceed until the amplitude becomes small enough. As we will
see later, the inflaton moves in a nearly circular orbit in the
complex plane, and so, the mass of the decay products does not change
during oscillations.  The decay into Higgs (Higgsino) becomes first
kinematically accessible when the amplitude of oscillations becomes
$\tilde \phi \sim \sqrt{\kappa}m_{3/2}/\lambda_\phi$.  If the inflaton
decayed at this epoch, the reheating temperature would be $T_R\sim
\kappa^{1/4}\lambda_\phi^{-1/2}m_{3/2}$. If the temperature estimated
this way exceeds the inflaton mass, the decay cannot complete the
reheating due to the back reaction of the thermal mass, and the actual
reheating temperature is given by $T_R \sim m_{\phi}$. This is indeed
the case if $\kappa \gtrsim \lambda_\phi^2$.  Otherwise, the reheating
temperature is given by $T_R\sim
\kappa^{1/4}\lambda_\phi^{-1/2}m_{3/2}$, but this does not modify our
argument qualitatively, because the $\kappa$ is expected to be not
much smaller than $O(\lambda_\phi^2)$ (see the discussion below
Eq.~(\ref{intK})).

In the above analysis we implicitly assumed that the inflaton is
oscillating about the origin with an amplitude smaller than
$\tilde{\phi}_{\rm e} = (m_{3/2}\kappa^{m/2}/\lambda)^{1/(m-1)}$ so
that the inflaton mass is simply given by $O(m_{3/2})$.  However, the
inflaton decay may proceed before the kination epoch ends, namely,
when the amplitude is larger than $\tilde{\phi}_{\rm e}$. In
particular, the inflaton has a larger mass (i.e. the energy per unit
inflaton quanta) in this regime, and so, it might be possible for the
inflaton to transfer a significant amount of energy to radiation.  The
effective mass for the inflaton during the kination regime is given by
$m_\phi (\tilde \phi) \sim \lambda {\tilde \phi}^{m-1}/\kappa^{m/2}$,
which decreases faster than the effective mass of $H_u$ and $H_d$.
Thus, if $m_\phi(\tilde \phi_{\rm b}) < \lambda_\phi \tilde \phi_{\rm
  b}/ \sqrt{\kappa}$, the inflaton decay does not occur during the
kination epoch, where $\tilde \phi_{\rm b}=\kappa^{n/(2n-2)}$ is the
inflaton field value at the beginning of kination.  This condition is
rewritten as $\kappa^{(m-n-1)/(2n-2)}\lambda < \lambda_\phi$.  If this
condition is not met, we need to take account of the radiation
generated during the kination epoch.  The radiation could dominate the
energy density of the Universe, and the inflaton may be subdominant at
the decay, or it may dominate the Universe again.

Furthermore, the actual situation could be more complicated, because
the inflaton also decays into the gauge bosons and gauginos through
the one-loop diagram, even if the tree-level decay is kinematically
forbidden.  Although the decay rate of such processes is suppressed, a
small amount of radiation produced by the process may dominate and
terminate the kination regime.  Thus, the details of the reheating
process and the resulting thermal history of the Universe depend much
on the model parameters.  In Sec.~\ref{sec4}, we fully take into
account the tree and one-loop decay processes.

\subsection{Preheating}

So far we have focused on the perturbative decay, but since the
inflaton starts to oscillate from the Planck scale, preheating could
be important especially for the early stage of the reheating epoch.
How fast and how long it proceeds depends on the interaction with the
Higgs field as well as the inflaton trajectory. In particular, it is
expected that the inflaton acquires a non-vanishing angular momentum
when it starts to oscillate because, the U(1) symmetry of the inflaton
potential around the origin is just an accidental one, and at large
scales, it is explicitly broken down to $Z_k$.  If the inflaton
acquires large enough angular momentum, it does not pass near the
origin, reducing the efficiency of the
preheating~\cite{Chacko:2002wr}. In fact we numerically followed the
inflaton dynamics and confirmed that the inflaton acquires a large
angular momentum at the end of inflation. See Fig.~\ref{fig:traj}. The
exception is the case of the $Z_{2n}$ symmetry, in which all $c_\ell$
with odd $\ell$ vanish; the inflationary trajectory coincides with the
real component of $\phi$, and so, no angular momentum is
induced. Therefore, in this case, preheating will play an important
role in the reheating.  We leave a detailed study of the preheating
for future work, and will focus on the case of $Z_k$ symmetry with $k
\leq n$.

\vspace{4mm}

We note that the angular momentum acquired by the inflaton is nothing
but a CP asymmetry of the inflaton. Therefore, if the inflaton has a
lepton- or baryon-number violating coupling like
\beq
W \;=\; \lambda^\prime \phi L H_u,
\eeq
where the $L$ is the lepton doublet, the asymmetry is transformed into
the lepton asymmetry.  Thus, with the aid of the sphaleron process
violating the $B+L$, the baryon asymmetry can be naturally produced by
the inflaton decay.

\begin{figure}[t]
\includegraphics[scale=0.6]{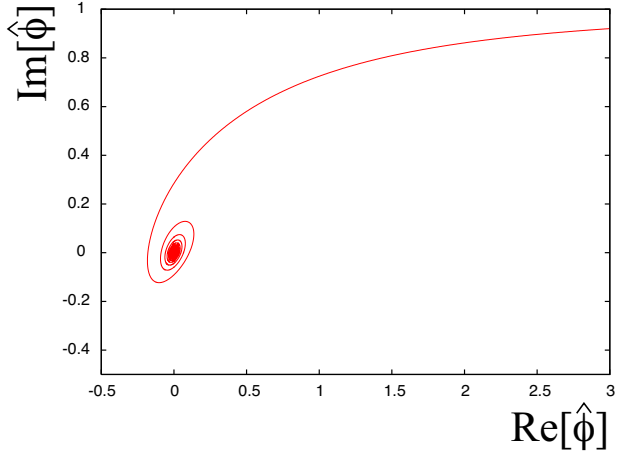}
\caption{ The inflationary trajectory for $n=m$ and $c_1 = -i$ in the
  ${\hat \phi} = \phi^2$ plane.  During inflation with $|{\hat \phi}|
  > 1$, $Im[\hat{\phi}]$ is stabilized at $1$.  After inflation the
  inflaton acquires a large angular momentum. }
\label{fig:traj}
\end{figure}

\section{Kination after inflation}
\label{sec3}

\subsection{Enhancement of gravity waves}
In the present model, the inflaton potential changes its form
depending on the field value $\phi$.  Correspondingly, the expansion
law, or the equation of state, of the Universe changes as well.  The
transition of the equation of state occurs as follows. After the
inflation ends at $|{\hat \phi}| \sim 1$, the inflaton oscillates in
the potential $V \propto |{\hat \phi}|^{2m/n}$ until the amplitude
becomes $|{\tilde \phi}| \sim \kappa^{1/(2n-2)}$. For the amplitude
smaller than $\kappa^{1/(2n-2)}$, the potential behaves as $V \propto
|{\tilde \phi}|^{2m}$: this is the kination regime.  As the amplitude
further decreases, the potential becomes finally dominated by the soft
mass term $\propto |{\tilde \phi}|^2$, which corresponds to the
matter-dominated Universe.  Then the inflaton decays and the radiation
dominant Universe begins.  The evolution of the Universe described
above leaves a characteristic imprint on the spectrum of inflationary
gravitational wave
background~\cite{Seto:2003kc,Boyle:2005se,Nakayama:2008ip,Mukohyama:2009zs}.
In particular, the kination after inflation enhances the gravity waves
at corresponding frequencies, and the direct detection of
gravitational waves by spaceborne laser interferometers may support or
refute the expansion history of the Universe.

In the following analysis we simply assume that the background plasma,
which may already exist in the kination regime, does not affect the
expansion history.  Let us denote the Hubble parameter at the
beginning and the end of kination as $H_{\rm b}$ and $H_{\rm e}$,
respectively.  They are estimated as $H_{\rm b} \sim \lambda
\kappa^{m/(2n-2)}$ and $H_{\rm e}\sim (\kappa^{m/2}
m_{3/2}^m/\lambda)^{1/(m-1)}$.  If the kination regime is followed by
the matter-dominated regime, the reheating temperature is given by
$T_R \sim m_{3/2}$, as explained before.  As an extreme case, we also
consider a case that inflaton decays as soon as the kination ends, and
the reheating temperature is then given by $T_R \sim \sqrt{H_{\rm
    e}}$, although some non-trivial process may be needed to realize
this situation.  During the kination stage, the scale factor grows as
$a(t) \propto t^{(m+1)/(3m)}$.  The spectrum of the gravitational
waves which enters the horizon at this stage looks like $\Omega_{\rm
  GW}(k)\propto k^{(2m-4)/(2m-1)}$ with a comoving wave number $k$.
Here we have defined
\begin{equation}
	\Omega_{\rm GW}(k)=\frac{1}{\rho_{0}}\frac{d\rho_{\rm GW}(k)}{d\ln k},
\end{equation}
where $\rho_0$ is the critical density of the Universe, and
$d\rho_{\rm GW}(k)/d\ln k$ denotes the energy density of the
gravitational waves per logarithmic frequency.

Therefore, the gravitational wave spectrum behaves as
\beq
\Omega_{\rm GW}(k) \propto \left \{
		\begin{array}{ll}
			k^{0} &{\rm for}~~k_{\rm eq}< k < k_{\rm R}, \\
			k^{-2} &{\rm for}~~k_{\rm R}< k < k_{\rm e}, \\
			k^{(2m-4)/(2m-1)} &{\rm for}~~k_{\rm e}< k < k_{\rm b}, 
		\end{array}
	\right.  \label{OmGW}
\eeq
if $k_{\rm R}< k_{\rm e}$, where $k_{\rm eq}, k_{\rm R}, k_{\rm e}$
and $k_{\rm b}$ denote the comoving horizon scale at the time of
matter-radiation equality, reheating, the end of kination and the
beginning of kination.  We find $k_{\rm R}= 2.6\times 10^{-5}~{\rm
  Hz}(T_{\rm R}/1{\rm TeV})(g_{*s}(T_{\rm R})/106.75)^{1/6}$, $k_{\rm
  e}=(H_{\rm e}/H_{\rm R})^{1/3}k_{\rm R}$ and $k_{\rm b}=(H_{\rm
  b}/H_{\rm e})^{(2m-1)/(3m)} k_{\rm e}$.  Importantly, the spectrum
is enhanced in the kination regime for $m>2$.

In the extreme case that the reheating occurs as soon as the kination
ends, we can set $k_{\rm R}=k_{\rm e}$ in Eq.~(\ref{OmGW}):
\beq
\Omega_{\rm GW}(k) \propto \left \{
		\begin{array}{ll}
			k^{0} &{\rm for}~~k_{\rm eq}< k < k_{\rm R}, \\
			k^{(2m-4)/(2m-1)} &{\rm for}~~k_{\rm R}< k < k_{\rm b}, 
		\end{array}
	\right.  \label{OmGW2}
\eeq
Thus the gravitational wave spectrum has no suppression in this case.

Fig.~\ref{fig:GW} shows the spectrum of gravitational waves for $n=4$
and $m=3$, for the case of (a) reheating immediately after kination
ends, (b) $\kappa = \lambda^2$ and (c) and $\kappa = \lambda$.  The
projected sensitivities of BBO/DECIGO~\cite{Harry:2006fi,Seto:2001qf}
its correlation analysis, and its ultimate sensitivities are also
shown.  We can clearly see the enhancement at high frequency
corresponding to the mode entering in the horizon during the kination
regime where the potential is given by $V \propto |{\tilde \phi}|^6$.
However, this enhancement is compensated to some extent by the
suppression due to the subsequent matter-dominated era.  Which effect
dominates depends on the parameter $\kappa$ and $\lambda$.

In most cases, the spectrum has characteristic features on its
magnitude and/or the tilt at observable frequencies.  If the
primordial gravitational waves is discovered by the future CMB
observations, we may have a good chance to detect them directly by the
planned space laser interferometers and to confirm the running kinetic
inflation model.

\begin{figure}[t]
\includegraphics[scale=0.7]{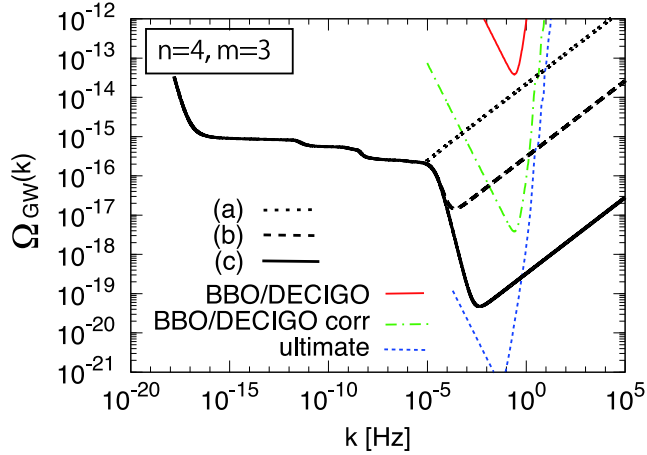}
\caption{ Spectrum of gravitational wave background for $n=4$ and
  $m=3$, for the case of (a) reheating immediately after kination
  ends, (b) $\kappa = \lambda^2$ and (c) and $\kappa = \lambda$.  The
  projected sensitivities of BBO/DECIGO and its correlation analysis
  are also shown.  }
\label{fig:GW}
\end{figure}

\subsection{Enhancement of unwanted relics : gravitino}

The modified expansion history of the Universe enhances not only
primordial gravitational waves, but also the abundance of unwanted
relics.  The gravitino is one of such dangerous relics. Let us
estimate its abundance.

At the inflationary stage, the Universe is regarded as being filled
with radiation with Hawking temperature $\sim H_I /(2\pi)$.  Thus
particles with masses smaller than $H_I$ would have thermal
abundances~\cite{Ford:1986sy}.  The gravitino number density at the
end of inflation is given by $n_{3/2}(H_I) \simeq \gamma H_I^3$, where
$\gamma \simeq 7\times 10^{-4}$.  The number density after the
reheating completes is estimated as
\begin{equation}
\begin{split}
	n_{3/2}(H_R)&=n_{3/2}(H_I)
	\left ( \frac{a(H_I)}{a(H_{\rm b})} \right )^3
	\left ( \frac{a(H_{\rm b})}{a(H_{\rm e})} \right )^3
	\left ( \frac{a(H_{\rm e})}{a(H_{\rm R})} \right )^3 \\
	&\sim
	\gamma H_I^3 \left ( \frac{H_{\rm b}}{H_I} \right )^{(m+n)/m}
	 \left ( \frac{H_{\rm e}}{H_{\rm b}} \right )^{(m+1)/m}
	 \left ( \frac{H_{\rm R}}{H_{\rm e}} \right )^{2}.
\end{split}
\end{equation}
Thus the gravitino abundance in terms of the number-to-entropy ratio is given by
\begin{equation}
	Y_{3/2}\equiv \frac{n_{3/2}}{s} \sim \frac{\gamma\lambda}{4}\frac{H_I}{M_P}
	\left ( \frac{H_I}{\lambda M_P} \right )^{(m-n)/m}.
\end{equation}
This is evaluated as
\begin{equation}
	Y_{3/2}\sim 7\times 10^{-14}\left( \frac{\lambda}{10^{-5}} \right)
	\left( \frac{H_I}{10^{14}{\rm GeV}} \right)
	\left ( \frac{H_I}{\lambda M_P} \right )^{(m-n)/m}.
\end{equation}
Here we have taken $T_R = m_{3/2}$.  The constraint reads
$m_{3/2}Y_{3/2}\lesssim 4\times 10^{-10}$GeV for a stable gravitino
and $Y_{3/2}\lesssim 10^{-16}$ for $m_{3/2} \sim 1$\,TeV and
$Y_{3/2}\lesssim 10^{-13}$-$10^{-12}$ for $m_{3/2}\sim 10$-$100$\,TeV
for an unstable gravitino~\cite{Kawasaki:2004qu}.  Therefore, even if
the gravity waves are enhanced by the kination epoch, the
gravitationally produced gravitino is cosmologically harmless for a
light stable gravitino ($m_{3/2}\lesssim 100$GeV) or a heavy gravitino
($m_{3/2}\gtrsim 10$TeV).  In the next section, we consider a case
that the inflaton is identified with a singlet field in the nMSSM,
where it turns out that there is no such enhancement, and therefore no
cosmological problem with the gravitationally produced gravitino.

\section{Inflaton embedded in nMSSM}
\label{sec4}

The inflaton and inflatino remain massless in the SUSY limit for $m
\geq 2$, as we have seen in Sec.~\ref{sec2}. If $n\geq 3$, once the
SUSY breaking is taken into account, the inflaton acquires a mass of
$O(m_{3/2})$, while the inflatino acquires its mass through a coupling
to the Higgs sector, (\ref{phi-higgs}). Since the inflaton and
inflatino are light, they may play an important role in the low
energy.

The MSSM plus a singlet that couples to the Higgs sector as
(\ref{phi-higgs}) is known as nMSSM. The main motivation to promote
the $\mu$-parameter to the singlet superfield is to solve the
$\mu$-problem. In order for the singlet to develop a VEV of the order
of the electroweak scale, a $Z_5$ or $Z_7$ R-symmetry is imposed on
the theory; the tadpole for the singlet is then radiatively induced at
sufficiently high order with a desired magnitude.  The difference from
the so called Next to Minimal SUSY SM (NMSSM) is the absence of both
the cubic self-coupling of the singlet and the domain wall problem
associated with the spontaneously broken $Z_3$ symmetry.  The discrete
R-symmetry adopted in nMSSM forbids the cubic self-coupling, and get
rid of the domain wall problem from the theory without destabilizing
the hierarchy.  The nMSSM has rich phenomenology; the singlino is a
viable candidate for dark matter, and the electroweak baryogenesis is
feasible~\cite{Huber:2006wf}. We here show that the running kinetic
inflation model can nicely fit with the nMSSM, with the additional
singlet being identified with the inflaton.

We have assigned non-zero discrete charges to $H_u H_d$ in
Eq.~(\ref{phi-higgs}), and so, we may wonder if domain walls are
produced at the electroweak phase transition. This can be solved if we
consider a discrete R-symmetry, which is explicitly broken by the
constant term $W_0$ in the superpotential. So far we have imposed two
symmetries, the R-symmetry and the discrete $Z_k$ symmetry on the
inflaton and $X$.  We can combine these two and consider a discrete
$Z_{k R}$ symmetry, with $k$ being a divisor of $2n$. Under the
$Z_{kR}$ symmetry, $(\phi^n-\phi^{\dag n})$ is invariant (up to a
phase factor for $k=2n$).  The charge assignment is given by
Table~\ref{tab:charge}.  Note here that the superpotential has a
R-charge $6$ in this charge assignment.

In order to have a successful inflation, there are several operators
that must be suppressed.  In particular, we require that $W \supset X,
X^2$ and $X^2\phi^m$ should be forbidden by the discrete R symmetry,
because these operators cannot be simply set to zero by the shift
symmetry.  We note that, although the discrete R symmetry on $\phi$ is
chosen so that it is consistent with the shift symmetry (\ref{sym}),
the operators including $\phi$ in the superpotential allowed by the
discrete R symmetry {\it do} break the shift symmetry.  Therefore we
cannot estimate naively the magnitudes of those operators. In this
sense, it is minimal to require that those three operators should be
forbidden by the discrete R symmetry.  We simply assume that the other
type of dangerous operators are sufficiently suppressed by the shift
symmetry. This restricts the possible values of $m$: only $m=1,3,6$
and $8$ are allowed for $m \leq 10$.  (In the case of the $Z_7$ R
symmetry, the allowed values of $m (\leq 14)$ are $m=1,2,3,4,8,9,10$
and $11$.)

In order to account for the correct abundance of the inflatino dark
matter, the effective coupling $\lambda_\phi/\sqrt{\kappa}$ must be of
order unity. This is indeed the case if $\kappa$ is radiatively
generated by the interaction (\ref{phi-higgs}).  Furthermore, the
inflatino must be long-lived to account for the dark matter. This is
the case if the inflatino is the LSP and the R-parity is conserved.
We note however that the fermionic superpartner of $X$ could be the
LSP, partially because we have forbidden its mass term $W \supset X^2$
by the discrete R symmetry.  Even in this case, the inflatino would be
sufficiently long-lived against the decay into $\tilde{X}$, if $m \geq
4$. Also, if $m=2$ and $\lambda/\kappa \approx
\lambda_\phi/\sqrt{\kappa}$, the inflatino mass can become comparable
to the ${\tilde X}$.

\begin{table}[htdp]
\begin{center}
\begin{tabular}{|c|c|c|c|c|c|c|}
\hline
     & $H_u$,\,$H_d$ & $Q$,\,$L$ & $U^c$,\,$D^c$,\,$E^c$ & $\phi$ & $X$ & $W$  \\  \hline
$Z_{5R}$, $Z_{7R}$ & $1$ & $2$ & $3$ & $4$ & $6-4m$ & $6$\\    \hline
\end{tabular}
\end{center}
\caption{The charge assignment of the $Z_{5R}$ and $Z_{7R}$ symmetries. 
  Note that the charge of the superpotential is 
  normalized to be $6$. }
\label{tab:charge}
\end{table}%

We have estimated the spectrum of primordial gravitational waves in
the setup embedded in the nMSSM, as shown in Fig.~\ref{fig:GW_nMSSM};
$n=5$ and $m=3$ with $\kappa =10^{-5}$ and $10^{-4}$ (left panel), and
$n=7$ and $m=4$ with $\kappa =10^{-6}$ and $10^{-5}$ (right panel).
As explained in Sec.~\ref{sec2}, the actual thermal history is
complicated due to the existence of the background plasma produced by
the loop-suppressed inflaton decay.  In this plot we have fully taken
into account the effect of decay products by the tree and one-loop
inflaton decay processes on the expansion history. The spectrum
becomes flat for $\kappa = 10^{-6}$ in the right panel of
Fig.~\ref{fig:GW_nMSSM}, because the radiation produced during the
kination comes to dominate the Universe for some time.

\begin{figure}[t]
\includegraphics[scale=0.6]{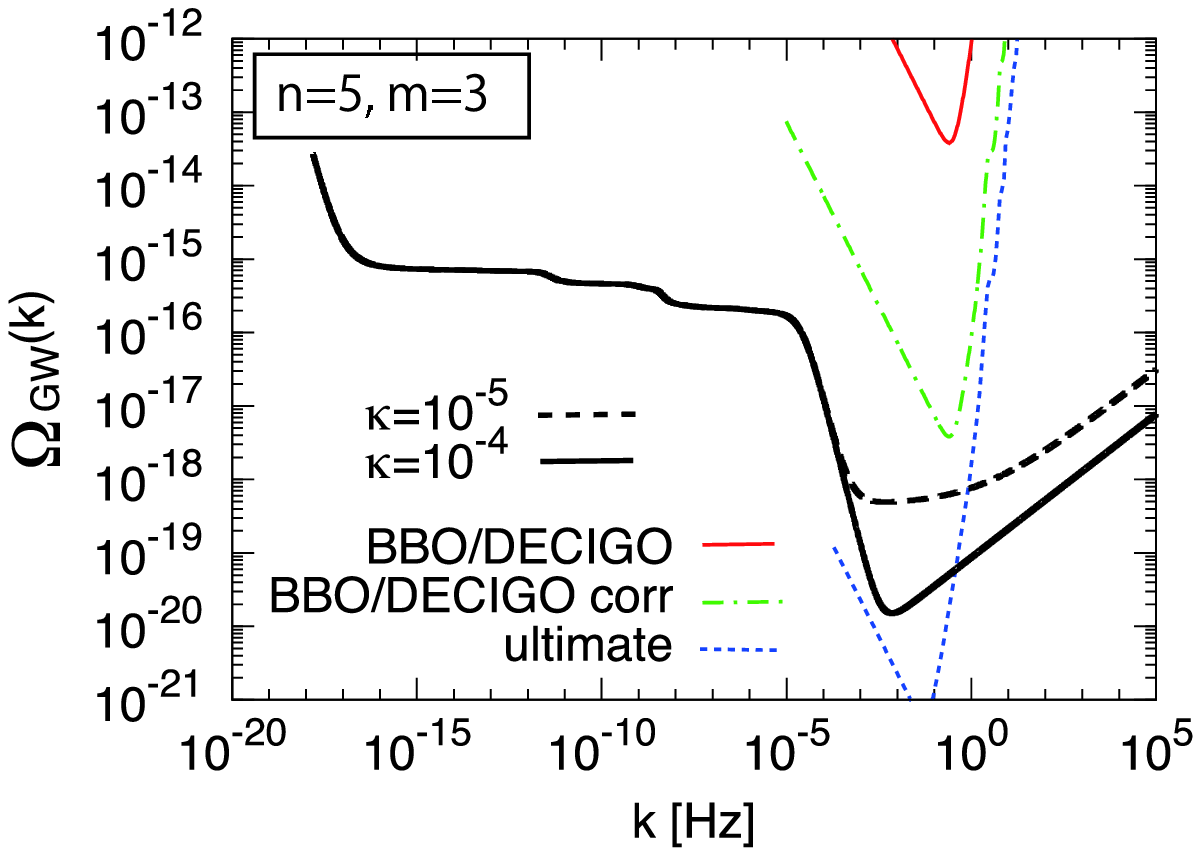}
\includegraphics[scale=0.6]{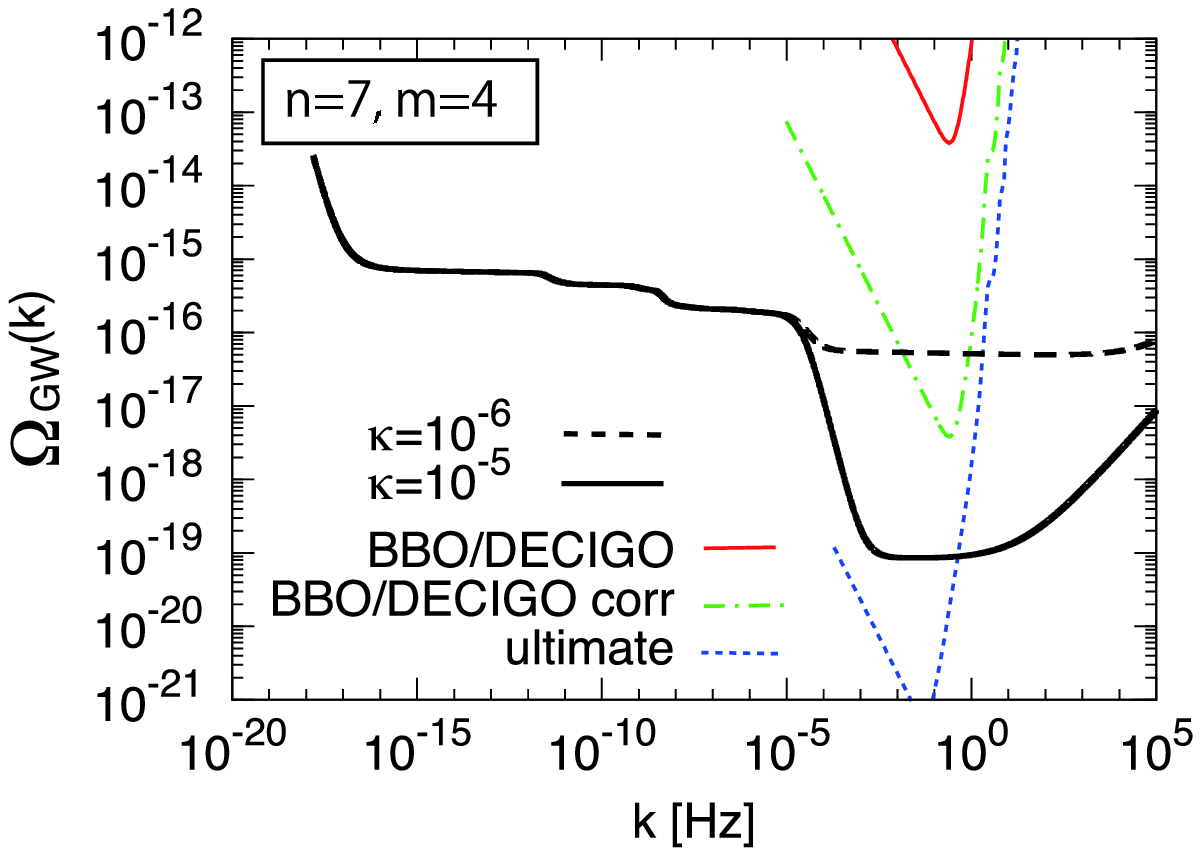}
\caption{ Spectrum of gravitational wave background for the case of
  $n=5$ and $m=3$ with $\kappa =10^{-5}$ and $10^{-4}$ (left panel),
  and $n=7$ and $m=4$ with $\kappa =10^{-6}$ and $10^{-5}$ (right
  panel), which are embedded in the nMSSM framework.  The projected
  sensitivities of BBO/DECIGO and its correlation analysis are also
  shown.  }
\label{fig:GW_nMSSM}
\end{figure}

\section{Conclusions}
\label{sec5}
In this paper we have studied the running kinetic inflation model,
motivated by a simple idea that the kinetic term may well
change its form if the inflaton moves over more than the Planck scale.
The growth of the kinetic term is actually advantageous for the inflation to occur,
because the effective potential for the canonically normalized field 
becomes flat. As a conrete example, we have constructed a variety of
chaotic inflation models in supergravity.
One of the features of the running kinetic inflation is that the power of the potential
increases after inflation: the potential is given by $\phi^{2m/n}$ during inflation, while 
it becomes $\phi^{2m}$ after inflation. The scalar spectral index and the tensor-to-scalar ratio are
given by $n_s = 1-\left(1+\frac{m}{n}\right) \frac{1}{N}$ and $r = \frac{8m}{n} \frac{1}{N}$, respectively.
Interestingly, the tensor-to-scalar ratio is generically large enough to be discovered by the CMB observation 
in future, especially by the Planck satellite. 

We have found that the model has a lot of interesting implications. In
particular, we have focused on the case of $m\geq 2$, in which the
inflaton is massless at the potential minimum in the SUSY limit. Once
the SUSY breaking effect is taken into account, the inflaton acquires
a mass of the order of the gravitino mass, in general. Because of such
a light mass, the thermal and non-thermal gravitino production can be
naturally avoided. Also we have seen that the gravity waves can be
enhanced due to the kination epoch which is naturally present in the
running kinetic inflation. Finally, we have pointed out that, in order
to have a successful reheating with such a light inflaton mass, the
inflaton needs to have unsuppressed couplings with the SM sector. One
plausible candidate is the coupling to the Higgs sector
(\ref{phi-higgs}). Then the model is similar to the so called nMSSM
with the singlet identified with the inflaton. The discrete R symmetry
introduced in nMSSM can be identified with the discrete symmetry on
the inflaton, and therefore the running kinetic inflation naturally
fits with nMSSM. The phenomenology of nMSSM is known to be rich;
the inflatino can account for the dark matter, and the inflaton coupling
enables the electroweak baryogenesis. 

The running inflation model, if realized in nature, will be confirmed
in coming years by the CMB observations. Interestingly, if $m
\geq 2$, the inflaton and/or inflatino may be discovered at the collider 
experiments such as LHC as well as the direct DM search.

\begin{acknowledgements}
  
  This work was supported by the Grant-in-Aid for Scientific Research
  on Innovative Areas (No. 21111006) [KN and FT] and Scientific
  Research (A) (No. 22244030) [FT], and JSPS Grant-in-Aid for Young
  Scientists (B) (No. 21740160) [FT].  This work was supported by
  World Premier International Center Initiative (WPI Program), MEXT,
  Japan.

\end{acknowledgements}



\end{document}